\documentclass[conference]{IEEEtran}
\usepackage{mathpazo}
\usepackage{times}

\usepackage{amsmath}
\usepackage{amsfonts}
\usepackage{latexsym}
\usepackage{amssymb}

\usepackage{upref}
\usepackage{theorem}
\usepackage{graphicx}
\usepackage{psfrag}
\usepackage{cite}

\theoremstyle{plain}
\theorembodyfont{\normalfont\slshape}

\newtheorem{thm}{Theorem$\!$}
\newenvironment{theorem}
{\begin{thm}\hspace*{-1ex}{\bf.}}{\end{thm}}

\newtheorem{clm}[thm]{Claim$\!$}

\newtheorem{lem}[thm]{Lemma$\!$}
\newenvironment{lemma}{\begin{lem}\hspace*{-1ex}{\bf.}}{\end{lem}}

\newtheorem{prop}[thm]{Proposition$\!$}

\newtheorem{cor}[thm]{Corollary$\!$}
\newenvironment{corollary}{\begin{cor}\hspace*{-1ex}{\bf.}}{\end{cor}}

\newtheorem{defn}[thm]{Definition$\!$}
\newenvironment{definition}{\begin{defn}\hspace*{-1ex}{\bf.}}{\end{defn}}

\newtheorem{xmpl}[thm]{Example$\!$}
\newenvironment{example}{\begin{xmpl}\hspace*{-1ex}{\bf.}}{\hfill$\Box$\end{xmpl}}

\newtheorem{cnstr}{Construction$\!$}
\newenvironment{construction}{\begin{cnstr}\hspace*{-1ex}{\bf.}}{\end{cnstr}}

\newcounter{enumrom}
\renewcommand{\theenumrom}{(\roman{enumrom})}

\makeatletter
\renewcommand{\@endtheorem}{\endtrivlist}
\makeatother

\makeatletter
\renewcommand{\thefigure}{{\@arabic\c@figure}}
\renewcommand{\fnum@figure}{{\bf Figure\,\thefigure}}
\makeatother

\newcommand{\cE}{\mathcal{E}}

\newcommand{\cG}{\mathcal{G}}
\newcommand{\cH}{\mathcal{H}}

\newcommand{\cM}{\mathcal{M}}

\newcommand{\mathset}[1]{\left\{#1\right\}}
\newcommand{\abs}[1]{\left|#1\right|}

\newcommand{\floorenv}[1]{\left\lfloor #1 \right\rfloor}
\newcommand{\parenv}[1]{\left( #1 \right)}

\newcommand{\be}[1]{\begin{equation}\label{#1}}
\newcommand{\ee}{\end{equation}}

\renewcommand{\leq}{\leqslant}

\renewcommand{\geq}{\geqslant}

\renewcommand{\Bbb}{\mathbb}

\newcommand{\Cref}[1]{Co\-ro\-lla\-ry\,\ref{#1}}

\renewcommand{\Bbb}{\mathbb}

\newcommand{\N}{{\Bbb N}}
\newcommand{\R}{{\Bbb R}}
\newcommand{\Z}{{\Bbb Z}}

\DeclareMathOperator{\GF}{GF}

\newcommand{\km}{k_{-}}
\newcommand{\kp}{k_{+}}
\newcommand{\qm}{q_{-}}
\newcommand{\qp}{q_{+}}
\newcommand{\rmm}{r_{-}}
\newcommand{\rp}{r_{+}}

\outer\def\proclaim #1. #2\par{\medbreak
 \noindent{\bf#1.\enspace}{\sl#2\par}%
 \ifdim\lastskip<\medskipamount \removelastskip\penalty55\medskip\fi}

\begin{document}

\IEEEoverridecommandlockouts 

\title{\textbf{Quasi-Cross Lattice Tilings \\ with Applications to Flash Memory}}

\author{
\IEEEauthorblockN{\textbf{Moshe Schwartz}}
\IEEEauthorblockA{Electrical and Computer Engineering\\
Ben-Gurion University of the Negev\\
Beer Sheva 84105, Israel\\
{\it schwartz@ee.bgu.ac.il}}
\thanks{
This work was supported in part by ISF grant 134/10.}
}

\maketitle


\begin{abstract}
We consider lattice tilings of $\R^n$ by a shape we call a
$(\kp,\km,n)$-quasi-cross.
Such lattices form perfect error-correcting codes which correct a single
limited-magnitude error with prescribed maximal-magnitudes of 
positive error and negative error (the ratio of which is called the
balance ratio). These codes can be used to correct both disturb and retention
errors in flash memories, which are characterized by having limited
magnitudes and different signs. 

We construct infinite families of
perfect codes for any rational balance ratio, and provide a specific
construction for $(2,1,n)$-quasi-cross lattice tiling. The constructions
are related to group splitting and modular $B_1$ sequences. We also
study bounds on the parameters of lattice-tilings by quasi-crosses, connecting
the arm lengths of the quasi-crosses and the dimension. We also prove
constraints on group splitting, a specific case of which shows that
the parameters of the lattice tiling of $(2,1,n)$-quasi-crosses is the
only ones possible.
\end{abstract}


\section{Introduction}
\label{sec:introduction}

Flash memory is perhaps the fastest growing memory technology
today. Flash memory cells use floating gate technology to store
information using trapped charge. By measuring the charge level in a
single flash memory cell and comparing it with a predetermined set of
threshold levels, the charge level is quantized to one of $q$ values,
conveniently chosen to be $\Z_q$. While originally $q$ was chosen to
be $2$, and each cell stored a single bit of information, current
\emph{multi-level flash} memory technology allows much larger values
of $q$, thus storing $\log_2 q$ bits of information in each
cell\footnote{%
It should be noted that other alternatives have been
  suggested to the conventional multi-level modulation scheme, such
  as, for example, rank modulation
  \cite{JiaMatSchBru09,JiaSchBru10,TamSch10,BarMaz10,WanBru10,Sch10,EngLanSchBru10}.
}.

As is usually the case, the stored charge levels in flash cells suffer
from noise which may affect the information retrieved from the cells.
Many off-the-shelf coding solutions exist and have been applied for
flash memory, see for example \cite{SunRosZha06,CheZhaWan08}. However,
the main problem with this approach is the fact that these codes are
not tailored for the specific errors occurring in flash memory and
thus are wasteful. A more accurate model of the flash memory channel
is therefore required to design better-suited codes.

The most notorious property of flash memory is its inherent asymmetry
between cell programming (charge injection into cells), and cell
erasure (charge removal from cells). While the former is easy to
perform on single cells, the latter works on large blocks of cells and
physically damages the cells. Thus, when attempting to reach a target
stored value in a cell, charge is slowly injected into the cell over
several iterations.  If the desired level has not been reached,
another round of charge injection is performed. If, however, the
desired charge level has been passed, there is no way to remove the
excess charge from the cell without erasing an entire block of
cells. In addition, the actions of cell programming and cell reading
disturb adjacent cells by injecting extra unwanted charge into them.
Because the careful iterative programming procedure employs small
charge-injection steps, it follows that over-programming errors, as
well as cell disturbs, are likely to have a small magnitude of error.

This motivated the application of the asymmetric limited-magnitude
error model to the case of flash memory
\cite{CasSchBohBru10,KloBosEla10}. In this model, a transmitted vector
$c\in\Z^n$ is received with error as $y=c+e\in\Z^n$, where we say that
$t$ asymmetric limited-magnitude errors occurred with magnitude at
most $k$ if the error vector $e=(e_1,\dots,e_n)\in\Z^n$ satisfies
$0\leq e_i \leq k$ for all $i$, and there are exactly $t$ non-zero
entries in $e$.  Not in the context of flash memory, it was shown in
\cite{AhlAydKhaTol04} how to construct optimal asymmetric
limited-magnitude errors correcting \emph{all} errors, i.e., $t$
equals the code length. General code constructions and bounds for
arbitrary $t$ were given in \cite{CasSchBohBru10}. More specifically,
for $t=1$, i.e., correcting a single error, codes were proposed in the
context of flash in \cite{KloBosEla10}, but were also described in the
context of semi-cross packing in the early work \cite{HicSte86}.

The main drawback of the asymmetric limited-magnitude error model is
the fact that not all error types were considered during the model
formulation. Another type of common error in flash memories is due to
\emph{retention} which is a slow process of charge leakage. Like
before, the magnitude of errors created by retention is limited,
however, unlike over-programming and cell disturbs, retention errors
are in the opposite direction.

We therefore suggest a generalization to the error model we call the
\emph{unbalanced limited-magnitude error model}. A transmitted vector
$c\in\Z^n$ is now received with error as the vector $y=c+e\in\Z^n$,
where we say that $t$ unbalanced limited-magnitude errors occurred if
the error vector $e=(e_1,\dots,e_n)\in\Z^n$ satisfies $-\km\leq e_i
\leq \kp$ for all $i$, and there are exactly $t$ non-zero entries in
$e$. Both $\kp$ and $\km$ are non-negative integers, where we call
$\kp$ the positive-error magnitude limit, and $\km$ the negative-error
magnitude limit.

\begin{figure}[ht]
\begin{center}
\includegraphics[scale=0.7]{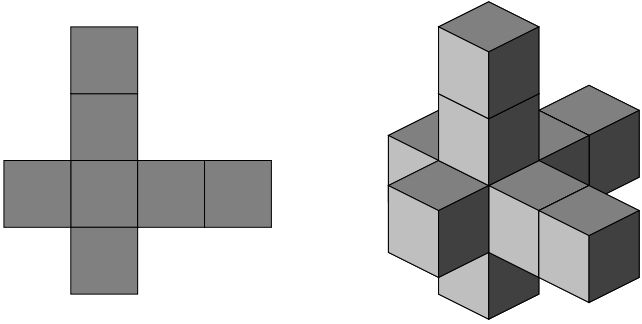}
\end{center}
\caption{A $(2,1,2)$-quasi-cross and a $(2,1,3)$-quasi-cross}
\label{fig:exampleqc}
\end{figure}

In this work we consider only single error-correcting codes. In
general, assuming at most a single error occurs, the error sphere
containing all possible received words $y=c+e$ forms a shape we call a
\emph{$(\kp,\km,n)$-quasi-cross} (see Figure \ref{fig:exampleqc}).
This is a generalization of the asymmetric semi-cross of
\cite{HicSte86,KloBosEla10} which we get when choosing $\km=0$, and
the full cross of \cite{Ste84} which we get when choosing
$\kp=\km$. To avoid these two studied cases we shall consider only $0
< \km < \kp$. An error-correcting code is a packing of pair-wise
disjoint quasi-crosses. We shall only consider perfect codes, i.e.,
tilings of the space, which form lattices, since these are easier to
analyze, construct, and encode, than non-lattice packings (see Figure
\ref{fig:latexample}).

The paper is organized as follows: In Section \ref{sec:prelim} we
introduce the notation and definitions used throughout the paper and
discuss connections with known results. We
continue in Section \ref{sec:bounds} with constructions of such tilings.
We follow in Section
\ref{sec:const} with simple bounds on the
parameter of lattice tilings of quasi crosses, and conclude
in Section \ref{sec:conc}.


\section{Preliminaries}
\label{sec:prelim}

\subsection{Quasi-Crosses, Tilings, and Lattices}
In the unbalanced limited-magnitude-error channel model, the transmitted (or
stored) word is a vector $v\in\Z^n$. A single error is a vector
in $e\in\Z^n$ all of whose entries are $0$ except for a single entry with
value belonging to the set
\[M=\mathset{-\km,\dots,-2,-1,1,2,\dots,\kp},\]
where the integers $0<\km<\kp$ are the negative-error and positive-error
magnitudes. For convenience we denote this set as $M=[-\km,\kp]^*$.
We denote $\beta=\km/\kp$ and call it the \emph{balance ratio}.
Obviously, $0 < \beta < 1$.

Given a transmitted vector $v\in\Z^n$, and provided at most
a single error occurred, the received word resides in the error sphere
centered about $v$ defined by
\[\cE(v)=\mathset{v}\cup \mathset{v+m\cdot e_i ~|~ i\in[n], m\in M},\]
where $[n]=\mathset{1,\dots,n}$, and $e_i$ denotes the all-zero vector
except for the $i$-th position which contains a $1$.
We call $\cE(0)$ a \emph{$(\kp,\km,n)$-quasi-cross}. By simple translation,
$\cE(v)=v+\cE(0)$ for all $v\in\Z^n$.

Following the notation of \cite{Ste84}, let
\[Q=\mathset{(x_1,\dots,x_n) ~|~ 0\leq x_i < 1, x_i\in\R}\]
denote the \emph{unit cube} centered at the origin. By abuse of terminology,
we shall also call the set of unit cubes $Q+\cE(v)$, a
$(\kp,\km,n)$-quasi-cross centered at $v$ for any $v\in\Z^n$. Examples
of such quasi-crosses are given in Figure \ref{fig:exampleqc}.
We note that the volume of $Q+\cE(v)$ does not depend on the choice
of $v$ and is equal to $n(\kp + \km)+1$.

A set $V=\mathset{v_1,v_2,\dots}\subseteq\Z^n$ defines a set of
quasi-crosses by simple translation:
$\mathset{\cE(v_1),\cE(v_2),\dots}$.  The set $V$ is said to be a
\emph{packing} of $\R^n$ by quasi-crosses if the translated
quasi-crosses are pairwise disjoint. The set $V$ is called a
\emph{tiling} if the union of the translated quasi-crosses equals
$\R^n$. If $V$ happens to be an additive subgroup of $\Z^n$ with a
basis $\mathset{b_1,b_2,\dots,b_n}$, then we call $V$ a
\emph{lattice}. The $n\times n$ integer matrix formed by placing the
elements of a basis as its rows is called a \emph{generating matrix} of the
lattice.

Let $\Lambda\subseteq\Z^n$ be a lattice with a generating matrix
$\cG(\Lambda)\in\Z^{n\times n}$ whose rows form a basis
$\mathset{b_1,b_2,\dots,b_n}\subseteq\Z^n$. A \emph{fundamental
  region} of $\Lambda$ is defined as
\[\mathset{\left.\sum_{i=1}^n \alpha_i b_i ~\right|~ \alpha_i\in\R, 0\leq \alpha_i < 1}.\]
It is easily seen, by definition, that $\Lambda$ tiles $\R^n$ with
translates of the fundamental region.

It is well known that the volume of a fundamental region does not
depend on the choice of basis for $\Lambda$ and equals $\det
\cG(\Lambda)$.  The \emph{density} of $\Lambda$ is defined as $1/\det
\cG(\Lambda)$ and if $\Lambda$ forms a packing of
$(\kp,\km,n)$-quasi-crosses, then the \emph{packing density} of
$\Lambda$ is defined as
\[\rho(\Lambda)=\frac{n(\kp + \km)+1}{\det \cG(\Lambda)},\]
which intuitively measures (for a large enough finite area) the ratio
of the area covered by $(\kp,\km,n)$-quasi-crosses centered at the
lattice points, to the total area. It follows that $0\leq
\rho(\Lambda)\leq 1$, and $\Lambda$ forms a tiling with
$(\kp,\km,n)$-quasi-crosses if and only if $\rho(\Lambda)=1$, i.e.,
$\det \cG(\Lambda)=n(\kp + \km)+1$.

\begin{figure}[ht]
\begin{center}
\includegraphics[scale=0.4]{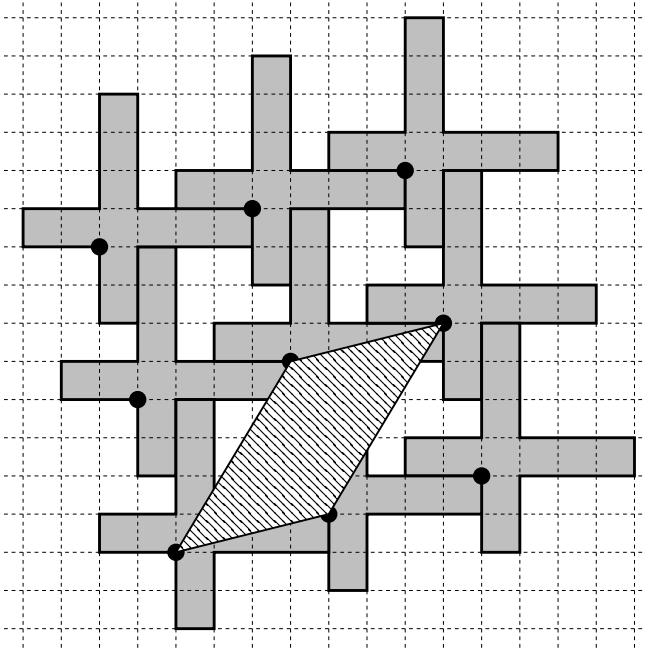}
\end{center}
\caption{Partial view of a lattice packing of a $(3,2,2)$-quasi-cross with
basis $b_1=(4,1)$, $b_2=(3,5)$, and
packing density $\frac{11}{17}$. Lattice points are marked with dots,
and the hatched area is a fundamental region.}
\label{fig:latexample}
\end{figure}

\begin{example}
\label{ex:lat}
If we take the $(3,2,2)$-quasi-cross, one can verify that the lattice $\Lambda$
with generating matrix
\[G(\Lambda)=\begin{pmatrix} 4 & 1 \\ 3 & 5 \end{pmatrix}\]
is indeed a lattice packing for this quasi-cross
(see Figure \ref{fig:latexample}). The resulting packing density
is
\[\rho(\Lambda)=\frac{2(3+2)+1}{\det \cG(\Lambda)}=\frac{11}{17}.\]
\end{example}

\subsection{Lattice Tiling via Group Splitting}

An equivalence between lattice packings and group splitting
was described in
\cite{Ste84,HicSte86}, which we describe here for completeness. Let
$G$ be an Abelian group, where we shall denote the group operation as
$+$. Given some $s\in G$ and a non-negative integer $m\in\Z$, we
denote by $ms$ the sum $s+s+\dots+s$, where $s$ appears in the sum $m$
times. The definition is extended in the natural way to negative
integers $m$.

A splitting of $G$ is a pair of sets,
$M\subseteq\Z\setminus\mathset{0}$, called the \emph{multiplier set},
and $S=\mathset{s_1,s_2,\dots,s_n}\subseteq G$, called the
\emph{splitter set}, such that the elements of the form $ms$, $m\in
M$, $s\in S$, are all distinct and non-zero in $G$. Next, we define a
homomorphism $\phi:\Z^n\rightarrow G$ by
\[\phi(x_1,x_2,\dots,x_n)=\sum_{i=1}^{n} x_i s_i.\]
If the multiplier set is $M=[-\km,\kp]^*$, then it may
be easily verifiable that $\ker \phi$ is a lattice packing of $\R^n$
by $(\kp,\km,n)$-quasi-crosses.
That $\ker \phi$ is a lattice is obvious. To show that the lattice is
a packing of $(\kp,\km,n)$-quasi-crosses, assume to the contrary two
such distinct quasi-crosses, $x=(x_1,\dots,x_n)$ and
$y=(y_1,\dots,y_n)$, have a non-empty intersection, i.e., $x+m_1
e_i=y+m_2 e_j$, where $m_1,m_2\in M$, then
\[m_1 s_i=\phi(x+m_1 e_i)=\phi(x+m_2 e_j)= m_2 s_j\]
which is possible only if $m_1=m_2$ and $i=j$, resulting in the two
quasi-crosses being the same one -- a contradiction.  The packing is a
tiling iff $\abs{G}=n(\kp+\km)+1$.

A simple representation of the lattice may also be given in matrix form:
Let $\cH=[s_1,s_2,\dots,s_n]$ be a $1\times n$ matrix over $G$. The
lattice $\Lambda$ is the set of vectors $x=(x_1,\dots,x_n)\in\Z^n$
such that $\cH x^T=0$. Thus, $\cH$ plays the role of a ``parity-check matrix''.

\begin{example}
\label{ex:parcheck}
Continuing Example \ref{ex:lat}, let $G=\Z_{17}$ and let
$M=\mathset{-2,-1,1,2,3}=[-2,3]^*$ stand for the multiplier set of the
$(3,2,n)$-quasi-cross. A possible splitting of $G$ is $S=\mathset{1,13}$, which
results in a parity-check matrix $\cH=[1,13]$ for the packing described
in Example \ref{ex:lat}.
\end{example}

Group splitting as a method for constructing error-correcting codes
was also discussed, for example, in the case of
shift-correcting codes \cite{Tam98} and integer codes \cite{Tam05}.

\subsection{Lattice Packings and Sequences}

It was noted in \cite{KloBosEla10} that there is a connection between
the codes suggested in \cite{KloBosEla10} (which are equivalent to
semi-cross packings) and a certain sub-case of sequences called
modular $B_h$ sequences. We detail the relevant connection in our case.

A $v$-modular $B_h(M)$ sequence, where
$M\subseteq\Z\setminus\mathset{0}$, is a subset\footnote{%
The actual sequence is the binary characteristic sequence of the subset to be
defined shortly.
} $S\subseteq\Z_v\setminus\mathset{0}$, whose
elements $S=\mathset{s_1,\dots,s_n}$ satisfy that all sums
$\sum_{i=1}^{h} m_i s_{i_j}$, where $1\leq i_1 < i_2 <\dots < i_h \leq n$,
and $m_i\in M$, are all distinct.

Thus, a $v$-modular $B_1(M)$ sequence is a splitting of $\Z_v$ defined
by $M$ and $S$. We note that a specific group is being split, i.e.,
a cyclic group.

As was also described in \cite{KloBosEla10}, when we have a
$v$-modular $B_1(M)$ sequence $S$, i.e., a splitting of $\Z_v$ by $M$
and $S$, and therefore a resulting $1\times n$ parity-check matrix
$\cH=[s_1,s_2,\dots,s_n]$, we can construct other packings, provided
the elements of $M$ are co-prime to $v$. This is
done by constructing any $k\times n(v^k-1)/(v-1)$ parity-check matrix
$\cH'$ containing all distinct column vectors whose top non-zero
element is from $S$. This is equivalent to a splitting of the non-cyclic
group $\Z_v^k$ by $M$ and $S$ being the columns of $\cH'$. We note
that if $\cH$ results in a tiling, then so does $\cH'$.

\section{Constructions for Tilings of Quasi-Crosses}
\label{sec:const}

We shall now consider constructions for lattice tilings of
$(\kp,\km,n)$-quasi-crosses. We first examine the case of a constant
balance ratio $\beta=\km/\kp$ and show that for any rational ratio
there exist infinitely-many tilings by splitting cyclic and non-cyclic
groups. We then focus on a particular case of $(2,1,n)$-quasi-crosses
and show an infinite family of tilings for them.

\subsection{Constant Balance-Ratio Quasi-Cross Tilings}

\begin{construction}
\label{con:gensplitcyc}
Let $0<\km<\kp$ be positive integers such that $\kp+\km=p-1$, where
$p$ is a prime. We set the multiplier set $M=[-\km,\kp]^*$.
Consider the cyclic group $G=\Z_{p^\ell}$, $\ell\in\N$. We split $G$ using a
splitter set $S$ constructed recursively in the following manner:
\begin{align*}
S_1 &= \mathset{1} \\
S_{i+1} &= p S_{i} \cup \mathset{\left. s\in\Z_{p^{i+1}} ~\right|~ s\equiv 1\pmod{p}}.
\end{align*}
The requested set is $S=S_\ell$.
\end{construction}
\begin{theorem}
The sets $S$ and $M$ from Construction \ref{con:gensplitcyc} split
$\Z_{p^\ell}$, forming a tiling of
$(\kp,\km,(p^\ell-1)/(p-1))$-quasi-crosses and a $p^\ell$-modular
$B_1(M)$ sequence.
\end{theorem}
\begin{IEEEproof}
The proof is by a simple induction. Obviously $M$ and $S_1=\mathset{1}$
split $\Z_p$. Now assume $M$ and $S_i$ split $\Z_{p^i}$. Let us consider
$M$, $S_{i+1}$, and $\Z_{p^{i+1}}$. We now show that if $ms=m's'$ in $\Z_{p^{i+1}}$,
$m,m'\in M$, $s,s'\in S_{i+1}$, then $m=m'$ and $s=s'$.

In the first case, given any $s\in S_{i+1}$, $p\nmid s$, and given
$m,m'\in M$, $m\neq m'$, since $M=[-\km,\kp]^*$, it follows that
$ms\neq m's$ since they leave different residues modulo $p$.  For the
second case, let $s,s'\in S$, $s'\neq s$, and let $m,m'\in M$, where
$m$ and $m'$ are not necessarily distinct. If $p|s'$ then $ms\neq
m's'$ since $p\nmid ms$ but $p|m's'$. We assume then that $s'\equiv
1\pmod{p}$. Write $s=qp+1$ and $s'=q'p+1$, $0\leq q,q'\leq p^{i}-1$,
then $ms=m's'$ implies $m=m'$ (by reduction modulo $p$). It then
follows that $mqp\equiv mq'p\pmod{p^{i+1}}$. But $\gcd(m,p)=1$ and so
$q\equiv q'\pmod{p^i}$, which (due to the range of $q$ and $q'$)
implies $q=q'$, i.e., $s=s'$.

For the last case, $s,s'\in pS_{i}$.
We note that the multiples of $p$ in $\Z_{p^{i+1}}$ are isomorphic to
$\Z_{p^i}$, and since $M$ and $S_i$ split $\Z_{p^i}$, for all
$m,m'\in M$, if $ms=m's'$ then $m=m'$ and $s=s'$.

Finally, $\abs{M}=p-1$, $\abs{S_{\ell}}=(p^\ell-1)/(p-1)$, and so
$\abs{M}\cdot\abs{S_{\ell}}+1=\abs{\Z_{p^\ell}}$, implying that the splitting
induces a tiling.
\end{IEEEproof}

The following construction splits a non-cyclic group of the same parameters.

\begin{construction}
\label{con:gensplitnoncyc}
Let $0<\km<\kp$ be positive integers such that $\kp+\km=p-1$, where
$p$ is a prime. We set the multiplier set $M=[-\km,\kp]^*$.
Consider the additive group of $G=\GF(p^\ell)$,
$\ell\in\N$. Let $\alpha\in \GF(p^\ell)$ be a primitive element,
and define $S=\mathset{P(\alpha) ~|~ P\in \cM^p_\ell[x]}$
where $\cM^p_\ell[x]$ denotes the set of all monic polynomials
of degree strictly less than $\ell-1$ over $\GF(p)$ in the indeterminate $x$.
\end{construction}

\begin{theorem}
The sets $S$ and $M$ from Construction \ref{con:gensplitnoncyc} split
the additive group of $\GF(p^\ell)$ and form a tiling of
$(\kp,\km,(p^\ell-1)/(p-1))$.
\end{theorem}
\begin{IEEEproof}
Since $\alpha$ is primitive in $\GF(p^\ell)$, the elements
$1,\alpha,\alpha^2,\dots,\alpha^{\ell-1}$ form a basis of the additive
group of $\GF(p^\ell)$ over $\GF(p)$. Since $M=\GF^*(p)$, it is easily
seen that $ms=m's'$, $m,m'\in M$, $s,s'\in S$, implies $m=m'$ and
$s=s'$. Again, by counting the size of $M$ and $S$, the splitting induces
a tiling.
\end{IEEEproof}

We point out several interesting observations. In Construction
\ref{con:gensplitnoncyc}, if we take $\ell=1$ we get $S=\mathset{1}$. For
$\ell > 1$, write then elements of $\GF(p^\ell)$ as length-$\ell$ vectors
over $\GF(p)$ (using the basis $1,\alpha,\dots,\alpha^{\ell-1}$,
with $\alpha$ a primitive element of $\GF(p^\ell)$). The elements of
$S$ then become the set of all vectors of length $\ell$ over $\GF(p)$
with the leading non-zero element being $1$. We will get the same set
by extending the ``matrix-extension'' method implied in \cite{KloBosEla10}
to our quasi-cross case.

Another interesting thing to note is that, using the same vector
notation as above, the parity-check matrix for the lattice is simply
the parity-check matrix of the $[\frac{p^\ell-1}{p-1},\frac{p^\ell-1}{p-1}-\ell,3]$
Hamming code over $\GF(p)$.

Yet another observation is that we can mix Constructions
\ref{con:gensplitcyc} and \ref{con:gensplitnoncyc}, by taking the
$p^\ell$-modular $B_1(M)$ sequence resulting from Construction
\ref{con:gensplitcyc} and applying the ``matrix'' method of
Construction \ref{con:gensplitnoncyc} to form a splitting of
$G=\Z_{p^\ell}\times\Z_{p^\ell}\times\dots\times\Z_{p^\ell}$ which
induces a tiling of quasi-crosses. The latter works since the elements
of $M$ are all co-prime to $p$.

Finally, as is shown in the next example, we observe that the lattice tilings
resulting from Constructions \ref{con:gensplitcyc} and \ref{con:gensplitnoncyc}
are not equivalent. Before we do so we need another definition. A lattice
$\Lambda\subseteq\Z^n$ has period $(t_1,\dots,t_n)\in\Z^n$ if whenever
$v\in\Lambda$, then also $v+t_i e_i\in\Lambda$ for all $i$. Lattices are
always periodic, and $t_i$ is the smallest positive integer for which
$t_i e_i\in\Lambda$.

\begin{example}
Consider six-dimensional lattice tilings of $(3,1,6)$-quasi-crosses. Using
Construction \ref{con:gensplitcyc} we construct a lattice
$\Lambda_1$ by splitting $\Z_{25}$ and getting a splitter set
$S=\mathset{1,5,6,11,16,21}$, resulting in a parity-check matrix
\[\cH_1=\begin{bmatrix} 1&5&6&11&16&21 \end{bmatrix} \]
over $\Z_{25}$. This produces a generating matrix for $\Lambda_1$
\[\cG_1=\begin{bmatrix}
25 & 0 & 0 & 0 & 0 & 0 \\
20 & 1 & 0 & 0 & 0 & 0 \\
19 & 0 & 1 & 0 & 0 & 0 \\
14 & 0 & 0 & 1 & 0 & 0 \\
 9 & 0 & 0 & 0 & 1 & 0 \\
 4 & 0 & 0 & 0 & 0 & 1
\end{bmatrix}.\]
We confirm that
\[\det \cG_1=25=6(3+1)+1\]
making $\Lambda_1$ a tiling for $(3,1,6)$-quasi-crosses.

If, on the other hand, we choose to use Construction \ref{con:gensplitnoncyc}
to construct a lattice $\Lambda_2$,
we split $\GF(5^2)$ to get a parity-check matrix
\[\cH_2=\begin{bmatrix}
0 & 1 & 1 & 1 & 1 & 1 \\
1 & 0 & 1 & 2 & 3 & 4
\end{bmatrix}\]
over $\GF(5)$.
A corresponding generating matrix is then
\[\cG_2=\begin{bmatrix}
5 & 0 & 0 & 0 & 0 & 0 \\
0 & 5 & 0 & 0 & 0 & 0 \\
4 & 4 & 1 & 0 & 0 & 0 \\
3 & 4 & 0 & 1 & 0 & 0 \\
2 & 4 & 0 & 0 & 1 & 0 \\
1 & 4 & 0 & 0 & 0 & 1 
\end{bmatrix}.\]
Again, we confirm $\det\cG_2=25$.

Finally, to show the lattices are not equivalent,
it is readily verified that the period of $\Lambda_1$ is
$(25, 5, 25, 25, 25, 25)$, while the period of $\Lambda_2$ is
$(5, 5, 5, 5, 5, 5)$.
\end{example}

The following shows there are infinitely-many tilings of quasi-crosses
of any given rational balance ratio.

\begin{theorem}
For any given rational balance ratio $\beta=\km/\kp$, $0<\beta<1$,
there exists an infinite sequence of quasi-crosses,
$\{(\kp^{(i)},\km^{(i)},n^{(i)})\}_{i=1}^\infty$, such that
$n^{(i)}<n^{(i+1)}$, $\km^{(i)}/\kp^{(i)}=\beta$, and
there exists a tiling of $(\kp^{(i)},\km^{(i)},n^{(i)})$-quasi-crosses,
for all $i\in\N$.
\end{theorem}
\begin{IEEEproof}
Given a rational $0<\beta<1$, let $\kp,\km\in\N$ be such that
$\km/\kp=\beta$. Denote $d=\kp+\km$ and consider the arithmetic
progression $1,1+d,1+2d,\dots,1+id,\dots$. Since $\gcd(1,d)=1$, by
Dirichlet's Theorem (see for example \cite{Apo76}), the sequence
contains infinitely-many prime numbers.  For any such prime, $p$,
there exists $q\in\N$ such that $q\kp+q\km=p-1$. We can then apply
Constructions \ref{con:gensplitcyc} and \ref{con:gensplitnoncyc} to
form tilings of $(q\kp,q\km,n)$-quasi-crosses with the required
balance ratio and $n$ unbounded.
\end{IEEEproof}

\subsection{Construction of $(2,1,n)$-Quasi-Cross Tilings}

We turn to constructing $(2,1,n)$-quasi-cross tilings and their
associated modular $B_1(M)$ sequences. The construction is similar in
flavor to Construction \ref{con:gensplitcyc}.

\begin{construction}
\label{con:21qc}
Let $\kp=2$, $\km=1$, and let the multiplier set be $M=\mathset{-1,1,2}$.
We split the group $G=\Z_{4^\ell}$, $\ell\in\N$, using a splitter set $S$
constructed recursively in the following manner:
\begin{align*}
S_1 &= \mathset{1} \\
S_{i+1} &= 4S_i \cup \mathset{s\in\Z_{4^{i+1}} ~|~ s\equiv 1\pmod{4}, 2s < 4^{i+1}}
\end{align*}
The requested set is $S=S_\ell$.
\end{construction}

\begin{theorem}
The sets $S$ and $M$ from Construction \ref{con:21qc} split $\Z_{4^\ell}$,
forming a tiling of $(2,1,(4^\ell-1)/3)$-quasi-crosses and a $4^\ell$-modular
$B_1(M)$ sequence.
\end{theorem}
\begin{IEEEproof}
The proof is by induction. The sets $M$ and $S_1$ obviously split $\Z_4$.
Assume $M$ and $S_i$ split $\Z_{4^i}$ and consider $M$ and $S_{i+1}$.
For convenience, denote
\[S'_{i+1}=\mathset{s\in\Z_{4^{i+1}} ~|~ s\equiv 1\pmod{4}, 2s < 4^{i+1}}.\]
It is easily seen that due to the restriction $2s<4^{i+1}$, the
elements of $S'_{i+1}$ and $-S'_{i+1}$ are distinct, and together they
contain all the odd integers in $\Z_{4^{i+1}}$. The elements of
$2S'_{i+1}$ are then also distinct and contain all the even integers
in $\Z_{4^{i+1}}$ leaving a residue of $2$ modulo $4$.

We are then left with all the multiples of $4$ in $\Z_{4^{i+1}}$ which form
a group isomorphic to $\Z_{4^i}$, and thus, by the induction hypothesis,
are split by $M$ and $4S_i$.

A simple counting argument shows that $\abs{M}=3$,
$\abs{S_\ell}=\frac{4^\ell-1}{3}$, and therefore
$\abs{M}\abs{S_\ell}+1=\abs{\Z_{4^\ell}}$. It follows that $M$ and $S_{\ell}$
split $\Z_{4^\ell}$ and form a tiling.
\end{IEEEproof}

We observe that in this case, since the elements of $M$ are not co-prime
to $4$, extending the matrix method from \cite{KloBosEla10} does not produce
a valid tiling or even packing. For example, if we were to take the
trivial $4$-modular $B_1(M)$ sequence, $\mathset{1}$ and attempt to create
a parity-check matrix over $\Z_4$
\[\cH=
\begin{bmatrix}
0 & 1 & 1 & 1 & 1 \\
1 & 0 & 1 & 2 & 3
\end{bmatrix}\]
we would find that $M$ together with the columns of $\cH$ is not a splitting
of $\Z_4^2$ since $2\cdot[1,0]^T=2\cdot[1,2]^T$ over $\Z_4$. Hence,
the lattice formed by the parity-check matrix $\cH$ is not a lattice
packing of $(2,1,5)$-quasi-crosses.

\section{Bounds on the Parameters of 
Lattice Tilings of Quasi-Crosses}
\label{sec:bounds}

In this section we focus on showing bounds on the parameters of
$(\kp,\km,n)$-quasi-cross tilings. We first consider the
restrictions $(\kp,\km,n)$-quasi-cross tilings imply on $\kp$, $\km$,
and $n$.  We then continue to study the group $G$ being split to
create the tilings, and show restrictions which, in particular, prove
that the parameters of the $(2,1,n)$-quasi-cross tiling of
Construction \ref{con:21qc} are unique.

\subsection{Dimension and Arm Length Bounds}

We first discuss bounds connecting the arm lengths of the quasi-cross and
the dimension of the tiling. Some of the theorems to follow may
be viewed as extensions to \cite{SteSza94}. 

\begin{theorem}
\label{th:nonexist}
For any $n\geq 2$, if
\[\frac{2\kp(\km+1)-\km^2}{\kp+\km}>n,\]
then there is no lattice tiling
of $(\kp,\km,n)$-quasi-crosses.
\end{theorem}
\begin{IEEEproof}
Given an integer $n\geq 2$, assume a $(\kp,\km,n)$-quasi-cross lattice tiling
$\Lambda$ exists.
Consider the plane $\mathset{(x,y,0,\dots,0) ~|~ x,y\in\Z}$.
Translates of this plane tile $\Z^n$. Within this plane, we look at
the subset
\begin{align*}
A=\{(x,y,0,\dots,0) ~|~ &0\leq x,y < \kp+2 \;\;\text{and}\\
&\text{$x<\km+2$ or $y<\km+2$}\}.
\end{align*}
It is easily seen that $A$ cannot contain two points from $\Lambda$,
or else the arms of two quasi-crosses overlap. Thus, the density of $\Lambda$
(which we know is exactly $1/(n(\kp+km)+1)$, since $\Lambda$ is a tiling)
cannot exceed the reciprocal of the volume of $A$, i.e.,
\[\frac{1}{n(\kp+\km)+1} \leq \frac{1}{(\kp+1)^2-(\kp-\km)^2}.\]
Rearranging gives us the desired result.
\end{IEEEproof}
\begin{corollary}
There is no lattice tiling of $\R^2$ by $(\kp,\km,2)$-quasi-crosses.
\end{corollary}
\begin{IEEEproof}
It is easily verifiable that for any $0<\km<\kp$,
\[\frac{2\kp(\km+1)-\km^2}{\kp+\km}>2.\]
\end{IEEEproof}

In the following theorem and corollary we can restrict the arm lengths of
quasi-crosses that lattice-tile $\R^n$.

\begin{theorem}
\label{th:smallarm}
For any $n\geq 2$, if a lattice tiling of $\R^n$ by
$(\kp,\km,n)$-quasi-crosses exists, then $\km\leq n-1$.
\end{theorem}
\begin{IEEEproof}
Let $0<\km<\kp$, and let $M=[-\km,\kp]^*$. Assume there is
a splitting of an Abelian group $G$ by $M$ and
$S=\mathset{s_1,\dots,s_n}$ which induces a lattice tiling of
$(\kp,\km,n)$-quasi-crosses, i.e.,
$\abs{G}=n(\kp+\km)+1$.

We first contend that for all $2\leq i\leq n$ there are integers
$x_i$ and $y_i$ such that
\begin{eqnarray*}
& \kp+1\leq x_i \leq \floorenv{\frac{n(\kp+\km)+1}{\km+1}} \\
& \abs{y_i} \leq \km \\
& s_1 x_i + s_i y_i = 0.
\end{eqnarray*}
To prove this, fix $i$ and let us look at the integers
\[0\leq a_1 \leq \floorenv{\frac{n(\kp+\km)+1}{\km+1}},
\qquad\qquad
0\leq a_2 \leq \km\]
and the sums $s_1 a_1 + s_i a_2$. Since
\begin{align*}
&\parenv{\floorenv{\frac{n(\kp+\km)+1}{\km+1}}+1}(\km+1) \geq \\
& \qquad\qquad \geq n(\kp+\km)+1-\km+\km+1 \\
& \qquad\qquad  =n(\kp+\km)+2 > \abs{G}
\end{align*}
by the pigeonhole principle there exist two distinct
pairs, $b_1,b_2$ and $c_1,c_2$, such that
\[s_1 b_1 + s_i b_2 = 0 \qquad \qquad s_1 c_1 + s_i c_2 = 0.\]
Assume w.l.o.g.~that $b_1\geq c_1$ and define
\[d_1=b_1-c_1 \qquad \qquad d_2=b_2-c_2.\]
We now get $s_1 d_1 + s_i d_2 =0$, where $(d_1,d_2)\neq (0,0)$. In addition,
\[0\leq d_1 \leq \floorenv{\frac{n(\kp+\km)+1}{\km+1}},
\qquad\qquad
\abs{d_2}\leq \km.\]
If $0\leq d_1\leq \kp$ then $s_1 d_1 = -s_i d_2$ contradicts the fact that
$S$ and $M$ split $G$. Thus,
\[\kp+1\leq d_1 \leq \floorenv{\frac{n(\kp+\km)+1}{\km+1}},\]
which proves our claim regarding the existence of $x_i$ and $y_i$.

For the rest of the proof we distinguish between two cases.
\textbf{Case 1:} There
exist $i\neq j$ such that $x_i=x_j$. In that case
\[0=s_1 x_i + s_i y_i = s_1 x_j + s_j y_j\]
in which case, $0=s_i y_i=s_j y_j$. However, $-\km\leq y_i,y_j\leq \km$
and to avoid contradicting the splitting, necessarily $y_i=y_j=0$.
It follows that $s_1 x_i = 0$. We now note that
\[-\km s_1,\dots,-s_1, 0, s_1,\dots,\kp s_1\]
are all distinct, and so the order of $s_1$ in $G$ is at least
$\kp+\km+1$, but has to divide $x_i$. Hence,
\[\kp+\km+1\leq x_i \leq \floorenv{\frac{n(\kp+\km)+1}{\km+1}}.\]
Rearranging the two sides gives us
\[\km\leq n-1-\frac{\km}{\kp+\km}\]
and since $0< \km < \kp$, necessarily $\km\leq n-2$.

\textbf{Case 2:} If $i\neq j$, then $x_i\neq x_j$. Thus, the number of
distinct values does not exceed their range, and we get
\[n-1 \leq \floorenv{\frac{n(\kp+\km)+1}{\km+1}}-\kp.\]
Rearranging this we get
\[\km\leq n-1+\frac{1}{\kp-1}.\]
If $\kp>2$ then, by the above, $\km\leq n-1$. If, however, $\kp=2$,
then $\km=1$ and obviously $\km\leq n-1$.
\end{IEEEproof}

\begin{corollary}
For any $n\geq 3$, if a lattice tiling of $\R^n$ by
$(\kp,\km,n)$-quasi-crosses exists and $\km>\frac{n}{2}-1$,
then
\[\kp\leq \begin{cases}
\frac{3n^2}{8} & \text{$n$ is even,} \\
\frac{3n^2-4n+1}{4} & \text{$n$ is odd.}
\end{cases}
\]
\end{corollary}
\begin{IEEEproof}
By Theorem \ref{th:nonexist}, a necessary condition for a lattice tiling
to exist is that
\[\frac{2\kp(\km+1)-\km^2}{\kp+\km}\leq n,\]
or after rearranging,
\[\kp(2(\km+1)-n)\leq \km^2 + n\km.\]
If $\km>\frac{n}{2}-1$, the left-hand side is positive and we get
\[\kp\leq \frac{\km^2 + n\km}{2(\km+1)-n}.\]
We need to maximize $\kp$, and by Theorem \ref{th:smallarm} we can
restrict ourselves to $\km\leq n-1$. The maximum is achieved at
$\km=\frac{n}{2}$ for $n$ even, and at $\km=\frac{n-1}{2}$ for $n$ odd.
Substituting back into the bound on $\kp$ gives the desired result.
\end{IEEEproof}

\subsection{Restrictions on the Split Group}

We now turn to examining connections between properties of the Abelian
group being split, $G$, and the multiplier and splitter sets, $M$ and $S$.
We shall eventually show, as a special case of the theorems presented, that
the $(2,1,n)$-quasi-cross tiles $\R^n$ only with the parameters of
Construction \ref{con:21qc}.
We follow the notation and definitions of \cite{SteSza94}.

\begin{definition}
Let $G$ be a finite Abelian group, and let $M$ and $S$ be the
multiplier and splitter sets forming a splitting of
$G$. We say the splitting is \emph{non-singular} if
$\gcd(\abs{G},m)=1$ for all $m\in M$. Otherwise, the splitting is
called \emph{singular}. If for any prime $p$ dividing the order of $G$
there is some $m\in M$ such that $p|m$, then the splitting is called
\emph{purely singular}.
\end{definition}

Given a finite $M\subseteq\Z$ and some prime $p\in\N$, we denote by
$\delta_p(M)$ the number of elements of $M$ divisible by $p$.
The following is an adaptation of \cite[p.~75, Corollary 2]{SteSza94}
for quasi-crosses, which is required for Theorem \ref{th:cyclic}.

\begin{lemma}
\label{lem:needforcyc}
Let $M=[-\km,\kp]^*$ be the multiplier set of the $(\kp,\km,n)$-quasi-cross.
Assume $M$ and $S$ are a purely-singular splitting of a finite Abelian
group $G$. Then $\delta_p(M)\geq \abs{M}/p^2$ for any prime divisor $p$
of $\abs{G}$.
\end{lemma}

\begin{IEEEproof}
Since the splitting is non-singular, for any prime divisor $p$ of $\abs{G}$,
$p$ divides some $m\in M=[-\km,\kp]^*$. Necessarily, $p\leq \kp$.
Let us assume
\[\km=\qm p + \rmm \qquad\qquad \kp=\qp p + \rp\]
where $0\leq \rmm,\rp < p$. We would like, therefore, to prove that
\[\delta_p(M)=\qp+\qm \geq \frac{\kp+\km}{p^2}.\]
After rearranging, this is equivalent to proving that
\[p\qp+p\qm\geq \frac{\rp+\rmm}{p-1}.\]
This obviously holds since $p\geq 2$, $\qp\geq 1$, and $\rp,\rmm\leq p-1$,
so
\[p\qp+p\qm\geq 2 \geq \frac{\rp+\rmm}{p-1},\]
proving the claim.
\end{IEEEproof}

Having proved Lemma \ref{lem:needforcyc}, the following theorem from
\cite{SteSza94} directly follows with the exact same proof.

\begin{theorem}\cite[p.~75, Theorem 9]{SteSza94}
\label{th:cyclic}
Let $M=[-\km,\kp]^*$ be the multiplier set of the $(\kp,\km,n)$-quasi-cross.
If $M$ splits $G$, then $M$ splits $\Z_{\abs{G}}$.
\end{theorem}

Theorem \ref{th:cyclic} is important since now, to show the existence or
nonexistence of a lattice tiling of $(\kp,\km,n)$-quasi-crosses, it is
sufficient to check splittings of $\Z_n$. We shall now do exactly that,
and reach the conclusion that $(2,1,n)$-quasi-crosses lattice-tile
$\R^n$ only with the parameters of Construction \ref{con:21qc}.

\begin{theorem}
\label{th:gcd}
Let $M=[-(k-1),k]^*$ be the multiplier set of the $(k,k-1,n)$-quasi-cross,
$k\geq 2$.
If $M$ splits a finite Abelian group $G$, $\abs{G}>1$,
then $\gcd(k,\abs{G})\neq 1$.
\end{theorem}
\begin{IEEEproof}
By Theorem \ref{th:cyclic} we may assume $G=\Z_q$. Denote the splitter
set $S=\mathset{s_1,s_2,\dots,s_n}$. It is easily seen that if
$\gcd(\ell,q)=1$, then $\ell S$ is also a splitter set. Since $1=m s$
for some $m\in M$ and $s\in S$, then $\gcd(m,q)=1$ and $1\in m S$. We can
therefore assume, w.l.o.g., that $s_1=1\in S$.

Since $M$ and $S$ split $\Z_q$, then $q\geq 2k$. If $q=2k$ the claim of the
theorem trivially holds. Assume then that $q>2k$. Let us consider the
unique factorization of $-k=ms_i$, $m\in M$ and $s_i\in S$. We note that
if $q>2k$, then $-k\not\equiv m\pmod{q}$ for all $m\in M$, and so
$s_i\neq s_1$.

If $-(k-1)\leq m\leq k-1$, then $-m\in M$ as well, and so
$k=-m s_i=k s_1$, and since $k\in M$, we get a contradiction to the splitting.
The only remaining option is that $m=k$, and $-k=ks_i$. If we
assume to the contrary that $\gcd(k,q)=1$, then
we can divide by $k$ and get $s_i=-1$. But then $-1=1\cdot s_i=(-1)\cdot s_1$,
where $1,-1\in M$, and we get a contradiction to the splitting again. It follows
that $\gcd(k,q)\neq 1$.
\end{IEEEproof}

\begin{corollary}
\label{cor:nonsing}
There is no non-singular splitting of $\Z_q$ by $M=[-(k-1),k]^*$.
\end{corollary}
\begin{IEEEproof}
Assume such a splitting exists, then $\gcd(q,m)=1$ for all $m\in M$, and
in particular $\gcd(q,k)=1$, contradicting Theorem \ref{th:gcd}.
\end{IEEEproof}

\begin{theorem}
\label{th:p2}
Let $M=[-2^w+1,2^w]^*$ be the multiplier set of the $(2^w,2^w-1,n)$-quasi-cross,
$w\in\N$. If $M$ splits $\Z_q$ then $q=2^{r(w+1)}$ for some $r\in\N$.
\end{theorem}
\begin{IEEEproof}
By Theorem \ref{th:gcd} and Corollary \ref{cor:nonsing}, $M$ cannot
split $\Z_q$ non-singularly and $\gcd(q,2^w)\neq 1$, i.e., $q$ is even.
Denote $q=t2^{r'}$, with $t,r'\in\N$, $t$ odd.

Let $S$ be the splitter set. Because of the splitting, every odd
number in $\Z_q$ is represented uniquely as $ms$, $m\in M$, $s\in S$, where
$m$ and $s$ are odd. There are $2^w$ odd numbers in $M$ and $t2^{r'-1}$ odd
numbers in $\Z_q$, so $2^w | t2^{r'-1}$ implying $r'\geq w+1$ and the
existence of exactly $t2^{r'-(w+1)}$ odd numbers in $S$.

Multiplying the odd numbers in $S$ by the elements of $M$ covers exactly
$2^{w-i}$ numbers in $\Z_q$ having a residue of $2^i$ modulo $2^{i+1}$, for 
all $0\leq i\leq w$.
The only, thus far, uncovered numbers in $\Z_q$ are those having
$0$ residue modulo $2^{w+1}$. These form a group isomorphic to
$\Z_{q/2^{w+1}}$. We also conclude that all even numbers in $S$ leave
a residue of $0$ modulo $2^{w+1}$.

We can therefore take $\Z_{q/2^{w+1}}$ and all the even numbers of $S$
divided by $2^{w+1}$ and repeat the argument above. We conclude
$q=t2^{r(w+1)}$ for some $r\in\N$. Also, the repetition of the above
argument repeatedly divides $q$ by $2^{w+1}$, and stops when we reach
the fact that $M$ splits $\Z_t$, $t$ odd. This is impossible by
Theorem \ref{th:gcd} unless $t=1$, which completes the proof.
\end{IEEEproof}

As a special case of the above theorems, we reach the following claim.

\begin{corollary}
The $(2,1,n)$-quasi-cross lattice-tiles $\R^n$ only with the parameters of
Construction \ref{con:21qc}.
\end{corollary}
\begin{IEEEproof}
Simply apply Theorem \ref{th:p2} with $w=1$ and compare with the parameters
of Construction \ref{con:21qc}.
\end{IEEEproof}

\section{Conclusion}
\label{sec:conc}

We considered lattice tilings of $\R^n$ by $(\kp,\km,n)$-quasi-crosses. These
lattices form perfect codes correcting a single error with limited magnitudes
$\kp$ and $\km$ for positive and negative errors, respectively. We have
seen how these lattice tilings are equivalent to certain group splittings,
and in certain cases (when the group is cyclic), to modular $B_1$ sequences.

We provided two constructions which may be used recursively to build
infinite families of such lattice tilings for any given rational balance
ration $\beta=\km/\kp$. We also specifically constructed an infinite
family of lattice tilings for the $(2,1,n)$-quasi-cross.

We followed by studying bounds on the parameters of such lattice tilings,
showing bounds connecting $\kp$, $\km$, and $n$. We also examined
restrictions on group splitting, and concluded through a special case of
the theorems presented, that $(2,1,n)$-quasi-crosses lattice-tile
$\R^n$ only with the parameters of the construction presented earlier.

We conclude with a computer search looking for lattice tilings of
$(\kp,\km,n)$-quasi-crosses. It was found that for all $0<\km<\kp\leq 10$
and split group $G=\Z_q$ of order $q\leq 100$, that only lattice tilings
with the parameters of the constructions provided in this paper exist.

\bibliographystyle{IEEEtranS}

\bibliography{allbib}

\end{document}